# A dynamic shim approach for correcting eddy current effects in diffusion-prepared MRI acquisition using a multi-coil AC/DC shim-array.

Congyu Liao[1,2], Jason P. Stockmann[3], Zhitao Li[2], Zhixing Wang[5], Mengze Gao[2], Lincoln Craven-Brightman[3], Monika Sliwiak[3], Charles Biggs[3], Jack Glad[4], Jiazheng Zhou[1], Yurui Qian[1], Zheng Zhong[2], Nan Wang[2,4], Hua Wu[6], Thomas Grafendorfer[7], Fraser Robb[7], Bernhard Gruber[3,8], Azma Mareyam[3], Adam B. Kerr[4,6], Xiaozhi Cao[2,4]*, and Kawin Setsompop[2,4]

[1] Department of Radiology and Biomedical imaging, University of California, San Francisco, USA

[2] Department of Radiology, Stanford University, Stanford, CA, USA

[3] Athinoula A. Martinos Center for Biomedical Imaging, Massachusetts General Hospital and Harvard Medical School, Charlestown, MA, USA

[4] Department of Electrical Engineering, Stanford University, Stanford, CA, USA

[5] Department of Radiation Oncology, University of Arkansas for Medical Sciences, Little Rock, AR, USA

[6] Stanford Center for Cognitive and Neurobiological Imaging, Stanford University, Stanford, CA, USA

[7] GE Healthcare, Milwaukee, WI, USA

[8] BARNLabs, Muenzkirchen, Austria

* Corresponding author
Xiaozhi Cao, PhD, xiaozhic@stanford.edu, Room 301, Packard Electrical Engineering Building, 350 Jane Stanford Way, Stanford, CA, 94305

# Abstract

**Purpose:** We developed a dynamic $B_0$ shimming approach using a 46-channel AC/DC shim array to correct phase errors caused by eddy currents from diffusion-encoding gradients in diffusion-prepared MRI, enabling high b-value imaging without the SNR loss from the use of magnitude stabilizer.

**Methods:** A 46-channel AC/DC shim array and corresponding amplifier system were built. Spin echo prescans with and without diffusion preparation were then used to rapidly measure eddy current induced phase differences. These phase maps were used as targets in an optimization framework to compute compensatory shim currents for multi-shot 3D diffusion-prepared acquisitions.

**Results:** The proposed method allows flexible use of the AC/DC shim array to correct undesirable eddy current effects in diffusion-prepared MRI. Phantom and *in vivo* experiments demonstrate whole-brain, cardiac-gated, multi-shot 3D diffusion-prepared imaging without the use of magnitude stabilizers. The approach enables preservation of full SNR while achieving reliable diffusion encoding at b-values up to 2000 s/mm$^2$.

**Conclusions:** This work demonstrates a new strategy for applying an AC/DC shim array to compensate for eddy current induced phase errors in diffusion-prepared MRI. By eliminating the need for magnitude stabilizer, it enables efficient high-quality diffusion imaging with full signal sensitivity retained.

# Introduction

Local multi-coil $B_0$ shimming (including "AC/DC" shim-arrays) [1–3] has proven to be effective in enhancing $B_0$ homogeneity and reducing echo-planar imaging (EPI) distortions in structural, functional, diffusion, and fetal MRI applications [4–10]. Conventional scanners are equipped with first-order (the linear gradients) and static second-order spherical harmonic shim coils, which generate a spatial magnetic profile to compensate for low-spatial order $B_0$ fields over the target volume. The AC/DC shim-arrays employing small shim coils patterned around the imaged object have been introduced as a convenient way to provide rapidly switchable, higher-order $B_0$ shim without the need to modify the MRI scanner substantially. One notable advantage of the AC/DC shim-array is its ability to rapidly switch shim currents without introducing image artifacts. This allows for the optimization of $B_0$ shimming on a slice-by-slice basis, leading to improved $B_0$ homogeneity and reduced geometric distortion in EPI acquisition. The dual-purpose AC/DC shim array provides high-spatial-order $B_0$ field control with high temporal resolution, enabling applications such as slice-by-slice dynamic shimming. In addition, the same array also serves as a high density receive array, offering good sensitivity and acceleration performance with parallel imaging [11,12].

Diffusion MRI plays a crucial role in numerous neuroscientific and clinical applications. In our previous work, we demonstrated the effectiveness of dynamic $B_0$ shimming using the AC/DC shim-array for significantly reducing local field inhomogeneity by more than 50%, as compared to the standard static second-order shimming, in diffusion-weighted spin-echo EPI acquisitions [7]. This dynamic shim approach can be further integrated with simultaneous multi-slice imaging [13–16], signal-to-noise ratio (SNR)-efficient EPI acquisition schemes [17–20], and advanced reconstruction techniques [21–27] to achieve comprehensive whole-brain coverage while minimizing image distortion. For instance, we have showcased the potential of combining dynamic shimming with multi-shot blip-up/down EPI acquisition and model-based reconstruction strategies [11]. This combination allows us to attain distortion-free, *in vivo* diffusion MRI at 600μm isotropic resolution, delivering high fidelity and sensitivity on a clinical 3T scanner. In the present study, in addition to addressing *in vivo* field inhomogeneity, we aim to harness the capabilities of the AC/DC shim-array for correcting other imperfections in MRI acquisitions, such as eddy currents and concomitant fields.

Strong diffusion-sensitizing gradients can induce eddy currents within the scanner bore, leading to spatiotemporal variations in both the magnitude and phase of the acquired signals. These variations depend on the strength and diffusion-direction of the diffusion-encoding gradients. The amplitude of the eddy-current response is influenced by the specific time constants and shows variability across scanners. There are numerous approaches to address eddy current-induced artifacts such as geometric distortions and blurring. These approaches include: (i) Temporal compensation of the input current waveform to the gradient coil, most notably with the pre-emphasis method[28]; (ii) The use of self-shielded gradient coil design[29,30] in minimizing eddy current interactions with surrounding structures has been well established in the literature[73,74]. These developments have played a critical role in enabling stable diffusion imaging. (iii) Post-processing techniques, such as the FSL 'eddy' tool[31,32], which corrects eddy current-induced distortions by leveraging the redundancy in directionally-sampled diffusion data along a sphere to derive a model of eddy current-induced magnetic fields up to the third order; and (iv) The use of a gradient impulse response function[33] or an NMR field camera[34,35] to measure eddy currents and subsequently incorporate these measurements into model-based reconstructions for eddy current correction[36,37]. Field camera-based corrections have been successfully implemented in various imaging techniques, including EPI, spiral imaging, and other complex k-space trajectories[36,38–40].

Diffusion-prepared (DP) acquisition is an alternative to the Stejskal-Tanner diffusion sequence, providing reduced readout-related blurring, and simultaneous encoding of $T_1$, $T_2$, and diffusion contrast for advanced multi-compartment modeling. Its benefits have been well established in prior work [42-43,64,75]. DP acquisitions employ a 90-degree tip-up pulse to convert diffusion-encoded transverse magnetization ($M_{xy}$) into longitudinal magnetization ($M_z$)[41]. This approach offers the advantage of high SNR with a short diffusion-preparation time and is compatible with various distortion-free readout methods, including fast-spin-echo[41] and steady-state sequences[42,43]. However, due to shot-to-shot phase variations resulting from physiological noise (e.g. cardiac pulsations) and eddy currents during diffusion encoding, these phase variations in $M_{xy}$ are subsequently stored as $M_z$ amplitude variations by the tip-up pulse. As such, the phase variations lead to undesirable magnitude variations following subsequent excitations and readouts. To address this issue, various methods have been proposed:

(i) A magnitude stabilizer is applied before the tip-up pulse to mitigate shot-to-shot variations[42]. However, this approach comes at the cost of losing up to half of the acquired signals because the magnitude stabilizer disperses transverse magnetization $M_{xy}$ before the tip-up pulse, leaving only the "net" $M_{xy}$ signals stored in longitudinal magnetization $M_z$. (ii) To avoid signal loss, techniques such as cardiac-gating and M1 or M2-compensated diffusion gradients[44–46] are used to reduce physiological-induced phase variations. Additionally, a pre-pulse method[47,48], which implements a large pulsed-gradient before the diffusion-preparation module, can generate significant eddy currents to correct eddy-current fields originating from diffusion gradients[48,69] or compensate eddy-current-induced phase offsets for diffusion-prepared acquisitions[47]. However, the pre-pulse method may not sufficiently compensate for eddy currents at high b-values with strong gradients. It is challenging to generate pre-pulses large enough to counteract the significant eddy currents induced by these strong gradients, thereby limiting its effectiveness in DP acquisition. In our work, we have taken a synergistic approach by using the AC/DC shim-array to correct eddy current-induced phase and employing M1-compensated diffusion-preparation with cardiac gating to correct physiological noise. Compared to a magnitude stabilizer, which results in the loss of half of the signal, this approach effectively minimizes shot-to-shot magnitude variations without sacrificing signal. It enables robust diffusion MRI with a short diffusion-preparation time while retaining the full signal level.

Similar to eddy currents, concomitant fields resulting from gradient encoding can introduce additional phase accrual during the acquisition[49], which results in image blurring and artifacts, particularly in acquisitions on high-performance gradient coil systems[50–52, 76] and low-field scanners[53], where such fields can be large. To address the issue of concomitant fields, a two-step approach is typically employed. First, the spatially varying concomitant fields are calculated using Maxwell's equations and the gradient coil settings. Subsequently, compensation is applied via sequence modification and/or further image reconstruction for phase error correction[54-57]. These methods have proven successful in compensating for concomitant fields on low-field scanners, as well as with both symmetric[56] and asymmetric[54,57] gradient coils. However, it is worth noting that the concomitant field effects can become much worse for slices offset along Z direction, and these compensation methods either require additional gradient blips with reduced scan efficiency[55,56], or are most effective for axial acquisitions with a specific positional offset along the z-axis only[57]. By incorporating concomitant field induced phase into the model-based reconstruction to correct for phase error accruals during the readout, the extra phase term perturbs the k-space image encoding, which makes it more ill-posed. Thus, it is still challenging to compensate complicated concomitant fields with oblique slice orientations positions or in simultaneous multi-slice acquisitions without the need of pulse sequence or reconstruction modification.

In this work, we propose the use of the AC/DC shim-array for eddy currents correction in diffusion-prepared MRI acquisitions by nulling the phase at the time of the tip-up pulse in a DP module. In addition to correcting eddy-current-induced phase errors, we also performed simulations to evaluate the potential of the AC/DC shim coil in compensating for concomitant field effects when using high-performance gradients. This work is an extension of our earlier work, which was reported as a conference abstract[71] and oral presentation in the Annual Meetings of International Society of Magnetic Resonance in Medicine (ISMRM) 2023.

## Methods

### AC/DC shim-array design

The design of the 46-channel AC/DC shim array used in this study follows the general concept of our previously developed 32-channel AC/DC shim array[2]. However, this new array was specifically designed and constructed for the 3T GE Ultra-High-Performance (UHP) scanner. The circuit architecture, cabling layout, helmet geometry, and amplifier interface were entirely redesigned to match the mechanical and electronic constraints of the UHP platform. Figure 1 illustrates the coil element loops, helmet, preamplifier circuit boards, shim connector socket, and shim amplifier enclosure. The coil includes 46 RF receive channels with 43 single loops configured to also carry DC (direct current) for $B_0$ shimming. One additional shim channel is provided by combining the two eye loops into a single shim current path, providing a total of 44 shim channels overall. The placement of small shim coils around the imaged object, along with the positioning of the shim amplifier beneath the patient table and routing of shim cables through the "caterpillar" cable guide, offers a convenient means to implement higher-order shimming without requiring significant modifications to the MRI scanner.

Calibrating the 46-channel AC/DC shim-array involves the following steps: (i) Inserting a large-field-of-view (FOV) phantom into the coil. (ii) Acquiring a $B_0$ mapping using a multi-echo gradient-echo sequence for each individual shim coil while switching the currents on a coil-by-coil basis. This step allows us to establish a $B_0$ map basis set for each shim coil, reflecting the sensitivity of field control for each individual coil (units of Hz/ampere). More details can be found in Ref.1. Figure 1(C) illustrates the calibration $B_0$ map basis set for all 44 shim channels for a single slice. (iii) Performing tests of RF transmission safety test to ensure there is no interference between the transmit RF volume coil and the AC/DC coils, no significant component heating on the AC/DC coil, as and no specific absorption rate (SAR) issues during scans. These steps are necessary only during the initial setup of the AC/DC shim-array. Further details on receive coil RF safety testing are available in Keil *et al*[68].

**Eddy currents correction using AC/DC shim-array in diffusion-prepared acquisition**

Figure 2(A) illustrates a typical pulse sequence diagram for DP acquisition where eddy current-induced phase variations can lead to signal loss following the tip-up pulse. Ideally, in the absence of eddy current-induced phase variations, all diffusion-encoded signals can be completely stored in longitudinal magnetization $M_z$ with the tip-up pulse. Subsequent to the DP module, consecutive excitations are implemented, and the stored diffusion-weighted signals are flipped down to the transverse plane and acquired until the next electrocardiogram-gating (ECG) trigger or pulse oximeter-gating (PG) trigger occurs. However, when there is a spatially varying phase error induced by eddy currents, the tip-up pulse may fail to flip all transverse magnetizations back to the longitudinal axis, resulting in signal loss. Figure 2(B) demonstrates signal loss in diffusion-preparation acquisition due to $\pi/2$ phase differences. It is important to note that if the phase error is close to $\pi/2$, it can result in 100% signal loss after the tip-up pulse. A magnitude stabilizer pulse can mitigate such magnitude variations at the cost of halving the acquired signal. Figure 2(C) presents signal evaluations with and without the magnitude stabilizer. Diffusion-prepared Fast Imaging with Steady-state Precession sequence (FISP) with constant flip-angles across 100 TRs at TR of 12ms is simulated using Bloch simulation. Given $T_1/T_2 = 800/60$ms, the transverse magnetization $M_{xy}$ for the 1st TR is 0.0816 without the magnitude stabilizer, while $M_{xy}$ for the 1st TR is 0.0408 with the magnitude stabilizer, indicating a 50% signal loss with the magnitude stabilizer in DP. Moreover, the subsequent TRs do not benefit from the $M_z$ recovery signal in the acquisition train as with magnitude stabilizer case that signal is crushed. As depicted in Figure 2(C), when a stabilizer is employed, the signal intensity rapidly decreases in the subsequent TRs. In previous studies, cardiac gating combined with M1-compensated diffusion gradients[47,77] have been shown to reduce phase variations caused by physiological noise. Additionally, pre-pulse gradients applied before the diffusion preparation (DP) module[47,48] have been used to mitigate eddy-current-induced phase errors. Nonetheless, its efficacy is not robust, especially at high b-values, as the pre-pulse method may not fully compensate for eddy currents induced by strong gradients. It is challenging to generate pre-pulses large enough to counteract significant eddy currents resulting from these gradients. In this work, we synergistically used the AC/DC shim-array to correct the eddy currents and M1-compensated diffusion-preparation with cardiac gating rather than a magnitude stabilizer to minimize physiological noise-induced shot-to-shot magnitude variations, enabling robust diffusion MRI with a short diffusion-preparation time and full signal level.

The eddy currents correction includes the following steps:

(i) A one-time eddy current-induced phase characterization: A spin-echo diffusion acquisition (Figure 3(A)) was implemented as a fast prescan to measure phase differences between non-diffusion (b=0) and diffusion-weighted acquisitions in a phantom, which corresponds to an estimation of the eddy current-induced phase in a diffusion prepared sequence, as shown in Figure 3(B). The phase differences were extracted by complex division between the diffusion-weighted and non-diffusion images. In cases where phase wrapping occurred in the phase difference maps, we applied the FSL PRELUDE function[31] to unwrap the phase and ensure accurate estimation. This prescan needed to be performed only once for each diffusion direction and gradient strength, rather than for every shot or readout.

(ii) The extracted phase difference from (i) was set as the input of DC-shim currents optimization. The optimal DC shim currents of the AC/DC shim-array were then computed using the calibration $B_0$ map basis set, to create opposite phase maps to compensate the eddy current-induced phase differences. Figure 4 shows the prescan measured phase difference between non-diffusion and diffusion acquisitions, and the predicted phase difference with shimming. With the shim optimization, the eddy current-induced phase differences were minimized to zero, which compensated eddy currents in DP acquisition. During the diffusion-encoding in the DP module, the shim duration was set to 5ms and the shim current amplitudes were scaled such that the overall amount of phase applied by the shim field was equal and opposite to the unwanted eddy-current induced phase at each voxel.

(iii) The calculated shim currents are applied during the diffusion-encodings of the diffusion-prepared sequence that has the same diffusion encodings as the prescan in (i). The shim currents update every TR through external triggers of the sequence (Figure 5), which enables compensation of differing eddy currents from different diffusion-directions. For *in vivo* diffusion acquisition, a PG trigger was implemented to minimize cardiac pulsation induced phase instability during the DP.

In addition to correcting eddy-current-induced phase errors, we also evaluated the potential of the AC/DC shim coil to compensate for concomitant field effects. As detailed in Supporting Information S2, our simulation results demonstrate the potential of using an AC/DC shim coil to compensate for concomitant field effects in double-oblique 2D spiral acquisitions with long readouts on a high-performance gradient system.

**Imaging protocols**

To validate our proposed method, we conducted both phantom and *in vivo* datasets acquisitions. In the case of the phantom scan, we initially performed a fast prescan acquisition on a short-$T_1$, head-shaped phantom without the tip-up pulse to characterize the eddy current-induced phase differences. A 3.4 mm isotropic low-resolution 3D dataset was acquired using a 64-shot spiral-projection trajectory covering the kx, ky, and kz axes (matrix size = 64 × 64 × 64). A single readout was applied after the diffusion-preparation (DP time = 50 ms, the same as in the *in vivo* acquisition), and the TR was set to 500 ms to prevent cross-contamination from eddy currents. Therefore, the acquisition time of the prescan was 0.5 × 64 = 32 s per b-value and per diffusion direction. We performed this single-acquisition prescan that included both a non-diffusion (b = 0) scan and diffusion-weighted scans at three b-values (b=600 s/mm$^2$, 1000 s/mm$^2$ and 2000 s/mm$^2$) and nine diffusion directions ([Gx, Gy, Gz]: [0.707, 0, 0.707], [-0.707, 0, 0.707], [0, 0.707, 0.707], [0, 0.707, -0.707], [0.707, 0.707, 0], [-0.707, 0.707, 0], [1, 0, 0], [0, 1, 0], [0, 0, 1]). The eddy-current-induced phase differences were

characterized from the phase difference between the diffusion-weighted and the non-diffusion scans. The prescan sequence diagram is depicted in Figure 3(A). Subsequently, using the measured phase differences, we calculated optimized shim currents to compensate for the phase differences. A least-square optimization with inequality constraints (total current per channel and total current for the whole array) was used to find shim currents which produce a spatial field profile that approximates the phase difference map, but with opposite polarity. Finally, to validate the proposed compensation method, we acquired phantom data using a diffusion-prepared 3D spiral projection sequence for three orthogonal diffusion directions ([Gx, Gy, Gz]: [1, 0, 0], [0, 1, 0], [0, 0, 1]) at a b-value of 600s/mm$^2$. In this acquisition, a single spiral interleave was acquired following each diffusion preparation, with TR and DP time set to 500 ms and 50 ms, respectively.

For the *in vivo* imaging, a cardiac-gated diffusion-prepared FISP (DP-FISP) sequence was implemented using 3D spiral-projection trajectories. A delay of 400 ms was applied after the PG to ensure that the acquisition is performed during end-diastole. The DP time was 50 ms, followed by 100 TRs of FISP acquisition. After these FISP readouts, data collection waited for the next PG pulse. The TR was approximately 2s per acquisition group, giving a total of 2 × 64 = 128 s per diffusion direction. For the *in vivo* diffusion tensor imaging (DTI) experiment, one b = 0 image and six diffusion-weighted images were acquired at both b = 600 s/mm$^2$ and 1000 s/mm$^2$ along the following directions: [Gx, Gy, Gz] = [0.707, 0, 0.707], [−0.707, 0, 0.707], [0, 0.707, 0.707], [0, 0.707, −0.707], [0.707, 0.707, 0], and [−0.707, 0.707, 0]. This results in a total scan time of 128 × 7 = 896s for whole-brain DTI at 1mm and 2mm isotropic resolution. Additionally, high-b-value *in vivo* data were acquired at 2 mm isotropic resolution for three diffusion directions ([Gx, Gy, Gz]: [1, 0, 0], [0, 1, 0], [0, 0, 1]) using pre-pulse diffusion gradients with a b-value of 2000 s/mm$^2$. To determine the optimal amplitude of the pre-pulse needed to compensate for the eddy-currents, the calibration spin-echo diffusion sequences were used to acquire images at b-values of 0 and 2000 s/mm$^2$. The acquisition with b = 0 (no eddy current from diffusion gradients) would be performed only once, whereas the acquisition with b = 2000 s/mm$^2$ would be repeated multiple times, each time with a different amplitude of the pre-pulse gradient. The pre-pulse gradients employed in the study were calibrated in three axes, ranging from -50 to 50 mT/m to minimize the impact of eddy currents in three diffusion directions. The optimal amplitude was then determined by finding the value that minimizes the phase difference between the b = 0 and 2000 s/mm$^2$ acquisitions. More details of pre-pulse implementation could be found in Ref.47.

All experiments were conducted using a 3T GE UHP scanner (GE Healthcare, Madison, WI, USA) and a 3T Siemens Prisma scanner (Siemens Healthineers, Erlangen, Germany). FOV-matched whole-brain B$_0$ maps (resolution: 2.0 ×2.0 ×2.0 mm$^3$) were obtained using a 7-second PhysiCal sequence[58] for fast *in vivo* B$_0$ calibration. To mitigate B$_0$-induced image blurring from the spiral readout, a multi-frequency interpolation (MFI) technique[59] was implemented with conjugate phase demodulation to achieve high image fidelity.

## Results

Figure 6 displays DP magnitude and phase images acquired from the head-shaped phantom, both with and without shim correction. Without eddy currents correction (first row in Figure 6), the bias caused by eddy current-induced phase differences leads to large signal loss, as indicated by the yellow arrows in the DP images. Conversely, the shim-corrected DP images exhibit higher signal intensities in these regions without any signal dropout. Furthermore, with eddy currents correction, the b=600 s/mm$^2$ DP images (second row in Figure 6) also show similar phase compared to the reference b=0 DP images (third row in Figure 6), demonstrating the eddy current-induced phase differences were compensated by shim currents.

Figure 7(A) presents whole-brain diffusion-weighted images with 2mm isotropic resolution acquired from six diffusion directions and Figure 7(B) shows corresponding fractional anisotropy (FA) maps at b=1000s/mm$^2$, both with and without shim correction. These FA maps were generated by fitting the data from six diffusion directions using the DTI model[60]. Without eddy currents correction, the phase differences induced by eddy currents lead to signal loss, as highlighted by the red arrows. However, the shim-corrected images showcase signal recovery in these regions, ultimately resulting in more accurate FA maps.

Figure 8 shows a representative bottom slice from a whole-brain DP dataset acquired at 1 mm isotropic resolution with six diffusion directions (b = 1000 s/mm$^2$) and shim correction. In addition to the diffusion-weighted images from each direction (Figure 8A), we also display several image frames sampled after the DP module (Figure 8B). These frames reflect the mixed $T_1$, $T_2$, and diffusion contrasts.

Figure 9 compares diffusion-prepared images acquired without correction, with pre-pulse correction, and with shim correction at b = 600 s/mm$^2$ across three orthogonal diffusion directions. As highlighted by the red arrows, both pre-pulse and shim corrections recovered regions of signal dropout in the brain.

Figure 10 presents diffusion-prepared images acquired at a higher b-value (b = 2000 s/mm$^2$) with and without shim correction and pre-pulse correction. While pre-pulse gradients alone were able to reduce some of the signal dropout caused by strong eddy currents, residual artifacts persisted in certain brain regions. In contrast, shim compensation provided further recovery in these areas. Taken together, Figures 8 and 9 demonstrate that the proposed shim compensation approach is effective at both low and high b-values. At b = 600 s/mm$^2$, shim correction performs comparably to pre-pulse methods, and at higher b-values, it offers additional advantages. Notably, shim compensation alone was sufficient to correct eddy-current-induced signal loss without the need for pre-pulse gradients, as shown in the fourth row of Figure 9. Although this approach required slightly higher current levels (increasing from 0.6A to 1A per channel), it eliminates the need for time-consuming pre-pulse calibration, thereby simplifying the experimental setup.

## Discussion

This study provides a demonstration of the applicability of the AC/DC shim-array for correcting undesirable eddy currents in diffusion MRI acquisitions. Through a series of phantom and *in vivo* experiments, we showcase its ability to enable high-fidelity, cardiac-gated, multi-shot 3D diffusion-prepared DTI acquisitions without the need for SNR-compromising magnitude stabilizers. Furthermore, we successfully achieved high-quality acquisitions at a high b-value of 2000 s/mm² by synergistically combining the shim-array with pre-pulse gradient mitigation strategies to minimize eddy current-induced phase errors in challenging scenarios. Our results highlight successful eddy current mitigation within a 3D DTI acquisition, paving the way for high-quality 3D *in vivo* diffusion MRI with high SNR. Additionally, our simulation study demonstrates an approach for resolving image blurring resulting from strong concomitant fields. This is particularly valuable in high-performance gradient systems and low-field MRI scanners. Simulation results further highlight the method's effectiveness in mitigating concomitant fields in non-Cartesian trajectories.

3D diffusion-prepared acquisitions with short TR and continuous readouts offer the advantages of reduced image distortion and improved SNR efficiency[42-43,64,75]. However, the 3D DP acquisition is sensitive to both physiological noise and eddy current-induced phase errors. These phase errors can lead to magnitude signal dropout following the tip-up pulse. This often leads to the common use of a magnitude stabilizer to achieve robust results. The magnitude stabilizer provides a correction for amplitude modulation caused by phase variations and ensures that only diffusion-prepared signal is captured, but this comes at the cost of a two-fold signal reduction. When the phase issue is corrected without a magnitude stabilizer, the signal loss is avoided, and additional Mz recovery signal is incorporated into the subsequent echo train. Whether the Mz recovery signal is desirable depends on the imaging goal: (1) If the goal is to obtain a purely diffusion-weighted signal, one may opt for sequences such as Turbo spin-echo (TSE) [64,65] or Gradient And Spin Echo (GRASE)[66] without variable flip angles, which are inherently sensitive only to the diffusion-prepared signal. This provides the benefit of full signal intensity while avoiding contamination from Mz recovery. (2) For diffusion-relaxometry imaging, the Mz recovery component can be beneficial. As demonstrated in our previous work on diffusion-prepared MR fingerprinting[47], retaining the Mz recovery enables joint modeling of $T_1$, $T_2$, and diffusion properties that could be beneficial for multi-tissue compartment mapping. In that study, we achieved significantly improved acquisition efficiency by acquiring the full signal without a magnitude stabilizer. Furthermore, we propose a more robust phase correction strategy that enables high-b-value diffusion-prepared acquisitions (e.g., b = 2000 s/mm²) without requiring a magnitude stabilizer, thereby allowing the full signal (including both the diffusion-prepared and Mz recovery components) to be utilized.

In this study, we mitigated physiological noise-induced phase errors, which are primarily caused by cardiac pulsations. This was achieved through the use of cardiac gating and M1 diffusion gradient compensation, as demonstrated by the reduced phase inconsistencies across shots. Additionally, we used the AC/DC shim array to address eddy-current-induced phase errors that cannot be fully corrected using the linear gradients alone. Eddy currents can introduce three types of imperfections: (i) accumulated eddy-currents-induced phase at the time of tip-up, which affects the tip-up efficiency, (ii) eddy currents during the tip-up pulse, and (iii) eddy currents after the tip-up pulse, which affect the readout and the stimulated echo in the steady-state sequence. Our proposed method addressed (i) by estimating and canceling the eddy-current-induced phase at the time of tip-up. In this study, a 90° adiabatic tip-up pulse was used to minimize (ii). As the FISP-based diffusion-prepared acquisition used in this study, smooth decay of the eddy-current field had minimal impact on the stimulated echo, and the highly segmented spiral further reduced residual effects from (iii).

While conventional eddy current nulling techniques[61–63] have been proposed to minimize eddy current fields *during* the EPI readout in diffusion imaging, they are not designed to minimize phase-accumulated from eddy current *across* the diffusion preparation period at the time point of the tip-up pulse. The introduction of pre-pulse correction, which uses eddy currents generated by pre-pulse gradients to compensate the eddy currents generated by diffusion encoding while maintaining a short diffusion-preparation time, has been successfully implemented in DTI-MR Fingerprinting study[47] with a b-value of 600 s/mm². The pre-pulse correction does not extend the acquisition time and DP time, especially since we are employing cardiac gating with a 400ms delay for *in vivo* study. However, this method still exhibits limitations in fully compensating for large eddy currents at high b-values, as shown in our experiments at b=2000 s/mm². In contrast, our proposed dynamic shim approach enables us to effectively compensate for eddy currents across different diffusion directions, even at high b-values. This overcomes the limitations associated with the pre-pulse correction method. Our proposed method could correct the eddy-current-induced phase at the time point of tip-up pulse rather than the instantaneous fields and longer shim pulses with lower currents can reproduce the required phase without exceeding hardware limits.

The eddy currents scale with the square of the slew rate. Therefore, when using head-only higher-performance scanners (e.g., Connectome 2.0[72] or MAGNUS[67]) with stronger diffusion gradients and higher slew rate, the resulting eddy currents will also be proportionally larger and may induce phase errors with higher spatial frequency content. Our 48-channel AC/DC shim array is capable of fitting such higher spatial frequency phase patterns. In particular, if the eddy-current-induced phase variation increases by 2x, the compensation can still be achieved either by (i) increasing the shim currents by 2x or (ii) extending the shim pulse duration by 2x. In our current implementation, we use a 5ms shim pulse, and the hardware supports up to 2 ampere per channel. Notably, we use less than 1 ampere per channel to correct for diffusion encoding with b = 2000 s/mm². This is well below the hardware current limit (2 ampere per channel), indicating that we can accommodate stronger corrections without the need to increase shim pulse duration.

In this study, we employed a calibration scan to directly measure the eddy current-induced phase error at the time point of tip-up and subsequently compensated for it using the AC/DC shim array. Notably, since eddy currents are subject-independent, the calibration scan was conducted using a phantom. On the whole-body scanners used in this study, we observed minimal eddy currents lasting longer than 50-100 ms. For such cases, the fast calibration scan implemented here was sufficient. In head-only high-performance gradient systems[67,72], however, substantially longer-lasting eddy currents have been reported[39]. In such scenarios, performing the calibration scan with a longer TR would be necessary to ensure that eddy currents from different slice acquisitions do not overlap. In this study, DP-FISP acquisition is performed with relatively long TR (~2 s) between diffusion encodings. Therefore, in our case, we expect minimal interaction between residual eddy currents and subsequent diffusion shots, especially on whole-body systems. It is also important to reduce calibration time when applying the proposed method to new diffusion protocols. Our results shown in Supporting Figure S1 demonstrate that if the diffusion mixing time is constant, the eddy-current-induced phase from any arbitrary diffusion direction can be effectively approximated as a linear combination of the eddy currents measured along the three orthogonal axes ($G_x$, $G_y$, and $G_z$). Additional implementation details are provided in Supporting Information S1.

To provide an initial demonstration of eddy current correction in the DP acquisition, we employed 3D spiral-projection readouts in DP-FISP sequence. This technique is not limited to a specific readout method and can be applied to other efficient sampling schemes, such as diffusion-prepared MR Fingerprinting[57], TSE[64,65] and GRASE[66] sequences. The use of a continuous echo-train following diffusion preparation can further improve the sampling efficiency of diffusion acquisitions, resulting in substantial reductions in acquisition time. Extending this approach to other acquisition types requires high temporal control of the shim waveform to accurately match the spatiotemporal evolution of the target fields. Such temporal control was not available in our current hardware at the time of the study but is under development.

The magnitude of the concomitant field is inversely proportional to the field strength and directly proportional to the square of gradient strength and off-center locations. This implies that concomitant fields can pose challenges not only in low-field MRI scanners, such as 0.55 T scanners but also in high-field MRI systems operating at 3T or 7T, especially those equipped with high-performance gradient systems[50–52,67], since the concomitant fields scale as a square of the gradient amplitude. Previous studies[55–57] have demonstrated several approaches to compensate for concomitant fields on low-field scanners, symmetric, and asymmetric gradient coils. However, these compensation methods either required extra time for added gradient blips[55] along with additional phase error correction during image reconstruction[56] or were only effective for acquisitions with a positional offset along the z-axis[57]. With our proposed dynamic updating method, we can effectively address the complex concomitant fields associated with oblique acquisition positions, without the sacrifice of scan efficiency. In future work, the concomitant field nulling method will be experimentally validated using upgraded shim amplifier hardware that can play out arbitrary current waveforms on every channel. Using arbitrary waveforms, the upgraded hardware could also be used for other applications such as joint shim-RF pulse design[10], multi-frequency wave-encoding[12], and multi-photon parallel-transmission[70]. Future research will also explore the application of this correction technique to more intricate readout k-space trajectories that can suffer from both large concomitant fields and gradient imperfections.

## Conclusion

In this study, we demonstrate a novel application of a custom-built 46-channel AC/DC shim array to dynamically correct eddy current–induced phase errors in challenging 3D diffusion-prepared MRI acquisitions. Our results show successful correction of these phase errors in multi-shot, cardiac-gated, 3D diffusion-prepared DTI, enabling acquisitions without the need for SNR-compromising magnitude stabilizers. This advancement has the potential to substantially improve the efficiency of 3D diffusion MRI while preserving high SNR.

## Acknowledgement

We thank Jeffrey Short for help with electronics assembly, Nicolas Arango for his expertise on shim amplifier debugging, Eugene Milshteyn and Arnaud Guidon for their support with running RF coil test scans on a GE scanner. In the preparation of this manuscript, the OpenAI's Large Language Model (LLM), specifically the GPT-4 architecture, was used for grammar check. This study is supported in part by NIH research grants: R01HD114719, R01MH116173, R01EB019437, U01EB025162, P41EB030006, R01EB033206, U24NS129893, U24EB028984.

## References

1. Stockmann JP, Wald LL. In vivo B0 field shimming methods for MRI at 7 T. *Neuroimage*. 2018;168:71-87. doi:10.1016/J.NEUROIMAGE.2017.06.013
2. Stockmann JP, Witzel T, Keil B, et al. A 32-channel combined RF and B0shim array for 3T brain imaging. *Magn Reson Med*. 2016;75(1):441-451. doi:10.1002/mrm.25587
3. Han H, Song AW, Truong TK. Integrated parallel reception, excitation, and shimming (iPRES). *Magn Reson Med*. 2013;70(1):241-247. doi:10.1002/mrm.24766
4. Juchem C, Umesh Rudrapatna S, Nixon TW, de Graaf RA. Dynamic multi-coil technique (DYNAMITE) shimming for echo-planar imaging of the human brain at 7 Tesla. *Neuroimage*. 2015;105:462-472. doi:10.1016/J.NEUROIMAGE.2014.11.011
5. Kim T, Lee Y, Zhao T, Hetherington HP, Pan JW. Gradient-echo EPI using a high-degree shim insert coil at 7 T: Implications for BOLD fMRI. *Magn Reson Med*. 2017;78(5):1734-1745. doi:10.1002/mrm.26563
6. Zhou J, Stockmann JP, Arango N, et al. An orthogonal shim coil for 3T brain imaging. *Magn Reson Med*. 2020;83(4):1499-1511. doi:10.1002/mrm.28010
7. Liao C, Stockmann J, Tian Q, et al. High-fidelity, high-isotropic-resolution diffusion imaging through gSlider acquisition with B1+ and T1 corrections and integrated ΔB0/Rx shim array. *Magn Reson Med*. 2020;83(1):56-67. doi:10.1002/mrm.27899
8. Overson DK, Darnell D, Robb F, Song AW, Truong TK. Flexible multi-purpose integrated RF/shim coil array for MRI and localized B0 shimming. *Magn Reson Med*. 2024;91(2):842-849. doi:10.1002/MRM.29891
9. Zhang M, Arango N, Stockmann JP, White J, Adalsteinsson E. Selective RF excitation designs enabled by time-varying spatially non-linear ΔB0 fields with applications in fetal MRI. *Magn Reson Med*. 2022;87(5):2161-2177. doi:10.1002/MRM.29114
10. Zhang M, Arango N, Arefeen Y, et al. Stochastic-offset-enhanced restricted slice excitation and 180° refocusing designs with spatially non-linear ΔB0 shim array fields. *Magn Reson Med*. 2023;90(6):2572-2591. doi:10.1002/MRM.29827
11. Liao C, Bilgic B, Tian Q, et al. Distortion-free, high-isotropic-resolution diffusion MRI with gSlider BUDA-EPI and multicoil dynamic B0 shimming. *Magn Reson Med*. 2021;86(2):791-803. doi:10.1002/mrm.28748
12. Xu J, Stockmann J, Bilgic B, et al. Multi-frequency wave-encoding (mf-wave) on gradients and multi-coil shim-array hardware for highly accelerated acquisition. In: *Int. Soc. Mag. Res. Med.* ; 2020:618. https://archive.ismrm.org/2020/0618.html. Accessed November 8, 2022.
13. Setsompop K, Gagoski BA, Polimeni JR, Witzel T, Wedeen VJ, Wald LL. Blipped-controlled aliasing in parallel imaging for simultaneous multislice echo planar imaging with reduced g-factor penalty. *Magn Reson Med*. 2012;67(5):1210-1224. doi:10.1002/mrm.23097
14. Ji Y, Hoge WS, Gagoski B, Westin CF, Rathi Y, Ning L. Accelerating joint relaxation-diffusion MRI by integrating time division multiplexing and simultaneous multi-slice (TDM-SMS) strategies. *Magn Reson Med*. 2022;87(6):2697-2709. doi:10.1002/MRM.29160
15. Cao X, Wang K, Liao C, et al. Efficient T2 mapping with blip-up/down EPI and gSlider-SMS (T2-BUDA-gSlider). *Magn Reson Med*. 2021;86(4):2064-2075. doi:10.1002/MRM.28872
16. Zhang Z, Cho J, Wang L, et al. Blip up-down acquisition for spin- and gradient-echo imaging (BUDA-SAGE) with self-supervised denoising enables efficient T2, T2*, para- and dia-magnetic susceptibility mapping. *Magn Reson Med*. April 2022. doi:10.1002/MRM.29219
17. Ji Y, Gagoski B, Hoge WS, Rathi Y, Ning L. Accelerated diffusion and relaxation-diffusion MRI using time-division multiplexing EPI. *Magn Reson Med*. 2021;86(5):2528-2541. doi:10.1002/MRM.28894
18. Liao C, Cao X, Cho J, Zhang Z, Setsompop K, Bilgic B. Highly efficient MRI through multi-shot echo planar imaging. In: Lu YM, Papadakis M, Van De Ville D, eds. *Wavelets and Sparsity XVIII*. Vol 11138. SPIE; 2019:43. doi:10.1117/12.2527183
19. Fair MJ, Liao C, Manhard MK, Setsompop K. Diffusion-PEPTIDE: Distortion- and blurring-free diffusion imaging with self-navigated motion-correction and relaxometry capabilities. *Magn Reson Med*. 2021;85(5):2417-2433. doi:10.1002/MRM.28579
20. Cho J, Liao C, Tian Q, et al. Highly accelerated EPI with wave encoding and multi-shot simultaneous

multislice imaging. *Magn Reson Med*. 2022;88(3):1180-1197. doi:10.1002/MRM.29291

21. Liao C, Yarach U, Cao X, et al. High-fidelity mesoscale in-vivo diffusion MRI through gSlider-BUDA and circular EPI with S-LORAKS reconstruction. *Neuroimage*. 2023;275:120168. doi:10.1016/J.NEUROIMAGE.2023.120168
22. Haldar JP, Liu Y, Liao C, Fan Q, Setsompop K. Fast submillimeter diffusion MRI using gSlider-SMS and SNR-enhancing joint reconstruction. *Magn Reson Med*. 2020;84(2):762-776. doi:10.1002/mrm.28172
23. Ramos-Llordén G, Ning L, Liao C, et al. High-fidelity, accelerated whole-brain submillimeter in vivo diffusion MRI using gSlider-spherical ridgelets (gSlider-SR). *Magn Reson Med*. March 2020:mrm.28232. doi:10.1002/mrm.28232
24. Bilgic B, Chatnuntawech I, Manhard MK, et al. Highly accelerated multishot echo planar imaging through synergistic machine learning and joint reconstruction. *Magn Reson Med*. May 2019:mrm.27813. doi:10.1002/mrm.27813
25. Chen Z, Liao C, Cao X, et al. 3D-EPI blip-up/down acquisition (BUDA) with CAIPI and joint Hankel structured low-rank reconstruction for rapid distortion-free high-resolution T2* mapping. *Magn Reson Med*. 2023;89(5):1961-1974. doi:10.1002/MRM.29578
26. Tian Q, Bilgic B, Fan Q, et al. DeepDTI: High-fidelity six-direction diffusion tensor imaging using deep learning. *Neuroimage*. 2020;219:117017. doi:10.1016/j.neuroimage.2020.117017
27. Liao C, Manhard MK, Bilgic B, et al. Phase-matched virtual coil reconstruction for highly accelerated diffusion echo-planar imaging. *Neuroimage*. 2019;194:291-302. doi:10.1016/j.neuroimage.2019.04.002
28. Ahn CB, Cho ZH. Analysis of the Eddy-Current Induced Artifacts and the Temporal Compensation in Nuclear Magnetic Resonance Imaging. *IEEE Trans Med Imaging*. 1991;10(1):47-52. doi:10.1109/42.75610
29. Trakic A, Liu F, Sanchez Lopez H, Wang H, Crozier S. Longitudinal gradient coil optimization in the presence of transient eddy currents. *Magn Reson Med*. 2007;57(6):1119-1130. doi:10.1002/MRM.21243
30. Spees WM, Buhl N, Sun P, Ackerman JJH, Neil JJ, Garbow JR. Quantification and Compensation of Eddy-Current-Induced Magnetic Field Gradients. *J Magn Reson*. 2011;212(1):116. doi:10.1016/J.JMR.2011.06.016
31. Jenkinson M, Beckmann CF, Behrens TEJ, Woolrich MW, Smith SM. Fsl. *Neuroimage*. 2012;62(2):782-790. doi:10.1016/j.neuroimage.2011.09.015
32. Alexander AL, Tsuruda JS, Parker DL. Elimination of eddy current artifacts in diffusion-weighted echo-planar images: The use of bipolar gradients. *Magn Reson Med*. 1997;38(6):1016-1021. doi:10.1002/mrm.1910380623
33. Campbell-Washburn AE, Xue H, Lederman RJ, Faranesh AZ, Hansen MS. Real-time distortion correction of spiral and echo planar images using the gradient system impulse response function. *Magn Reson Med*. 2016;75(6):2278-2285. doi:10.1002/MRM.25788
34. Vannesjo SJ, Haeberlin M, Kasper L, et al. Gradient system characterization by impulse response measurements with a dynamic field camera. *Magn Reson Med*. 2013;69(2):583-593. doi:10.1002/MRM.24263
35. Gross S, Barmet C, Dietrich BE, Brunner DO, Schmid T, Pruessmann KP. Dynamic nuclear magnetic resonance field sensing with part-per-trillion resolution. *Nat Commun*. 2016;7(1):1-7. doi:10.1038/ncomms13702
36. Lee Y, Wilm BJ, Brunner DO, et al. On the signal-to-noise ratio benefit of spiral acquisition in diffusion MRI. *Magn Reson Med*. 2021;85(4):1924-1937. doi:10.1002/mrm.28554
37. Dubovan PI, Allan Baron C, Corey C, Baron A. Model-based determination of the synchronization delay between MRI and trajectory data. *Magn Reson Med*. 2023;89(2):721-728. doi:10.1002/MRM.29460
38. Wilm BJ, Hennel F, Roesler MB, Weiger M, Pruessmann KP. Minimizing the echo time in diffusion imaging using spiral readouts and a head gradient system. *Magn Reson Med*. June 2020:mrm.28346. doi:10.1002/mrm.28346
39. Ramos-Llordén G, Park DJ, Kirsch JE, et al. Eddy current-induced artifact correction in high b-value

ex vivo human brain diffusion MRI with dynamic field monitoring. *Magn Reson Med*. 2024;91(2):541-557. doi:10.1002/MRM.29873

40. Dubovan PI, Gilbert KM, Baron CA. A correction algorithm for improved magnetic field monitoring with distal field probes. *Magn Reson Med*. 2023;90(6):2242-2260. doi:10.1002/MRM.29781
41. Van AT, Cervantes B, Kooijman H, Karampinos DC. Analysis of phase error effects in multishot diffusion-prepared turbo spin echo imaging. *Quant Imaging Med Surg*. 2017;7(2):238-250. doi:10.21037/qims.2017.04.01
42. Gao Y, Han F, Zhou Z, et al. Multishot diffusion-prepared magnitude-stabilized balanced steady-state free precession sequence for distortion-free diffusion imaging. *Magn Reson Med*. 2019;81(4):2374-2384. doi:10.1002/mrm.27565
43. Lu L, Erokwu B, Lee G, et al. Diffusion-prepared fast imaging with steady-state free precession (DP-FISP): A rapid diffusion MRI technique at 7 T. *Magn Reson Med*. 2012;68(3):868-873. doi:10.1002/mrm.23287
44. Nguyen C, Fan Z, Xie Y, et al. In vivo Diffusion-Tensor MRI of the Human Heart On a 3T Clinical Scanner: An Optimized Second Order (M2) Motion Compensated Diffusion-Preparation. *Magn Reson Med*. 2016;76(5):1354. doi:10.1002/MRM.26380
45. Kara D, Koenig K, Lowe M, Nguyen CT, Sakaie K. Facilitating diffusion tensor imaging of the brain during continuous gross head motion with first and second order motion compensating diffusion gradients. *Magn Reson Med*. 2023. doi:10.1002/MRM.29924
46. Nguyen C, Fan Z, Sharif B, et al. In vivo three-dimensional high resolution cardiac diffusion-weighted MRI: A motion compensated diffusion-prepared balanced steady-state free precession approach. *Magn Reson Med*. 2014;72(5):1257-1267. doi:10.1002/MRM.25038
47. Cao X, Liao C, Zhou Z, et al. DTI-MR fingerprinting for rapid high-resolution whole-brain T1, T2, proton density, ADC, and fractional anisotropy mapping. *Magn Reson Med*. 2023. doi:10.1002/MRM.29916
48. Alexander A , et al. Elimination of eddy current artifacts in diffusion-weighted echo-planar images: the use of bipolar gradients. Magnetic Resonance in Medicine. doi:10.1002/mrm.1910380623
49. Du YP, Zhou XJ, Bernstein MA. Correction of concomitant magnetic field-induced image artifacts in nonaxial echo-planar imaging. *Magn Reson Med*. 2002;48(3):509-515. doi:10.1002/mrm.10249
50. Huang SY, Witzel T, Keil B, et al. Connectome 2.0: Developing the next-generation ultra-high gradient strength human MRI scanner for bridging studies of the micro-, meso- and macro-connectome. *Neuroimage*. 2021;243:118530. doi:10.1016/J.NEUROIMAGE.2021.118530
51. Feinberg DA, Beckett AJS, Vu AT, et al. Next-generation MRI scanner designed for ultra-high-resolution human brain imaging at 7 Tesla. *Nature Methods 2023 20:12*. 2023;20(12):2048-2057. doi:10.1038/s41592-023-02068-7
52. Foo TKF, Tan ET, Vermilyea ME, et al. Highly efficient head-only magnetic field insert gradient coil for achieving simultaneous high gradient amplitude and slew rate at 3.0T (MAGNUS) for brain microstructure imaging. *Magn Reson Med*. 2020;83(6):2356-2369. doi:10.1002/MRM.28087
53. Shetty AS, Ludwig DR, Ippolito JE, Andrews TJ, Narra VR, Fraum TJ. Low-Field-Strength Body MRI: Challenges and Opportunities at 0.55 T. *Radiographics*. 2023;43(12). doi:10.1148/RG.230073/ASSET/IMAGES/LARGE/RG.230073.FIG22.JPEG
54. Abad N, Lee SK, Ajala A, et al. Calibration of concomitant field offsets using phase contrast MRI for asymmetric gradient coils. *Magn Reson Med*. 2023;89(1):262-275. doi:10.1002/MRM.29452
55. Wang Z, Ramasawmy R, Feng X, Campbell-Washburn AE, Mugler JP, Meyer CH. Concomitant magnetic-field compensation for 2D spiral-ring turbo spin-echo imaging at 0.55T and 1.5T. *Magn Reson Med*. 2023;90(2):552-568. doi:10.1002/MRM.29663
56. Cheng JY, Santos JM, Pauly JM. Fast concomitant gradient field and field inhomogeneity correction for spiral cardiac imaging. *Magn Reson Med*. 2011;66(2):390-401. doi:10.1002/mrm.22802
57. Weavers PT, Tao S, Trzasko JD, et al. B0 concomitant field compensation for MRI systems employing asymmetric transverse gradient coils. *Magn Reson Med*. 2018;79(3):1538-1544. doi:10.1002/MRM.26790/ASSET/SUPINFO/MRM26790-SUP-0001-SUPPINFO01.DOCX
58. Iyer S, Liao C, Li Q, … MMP of the, 2020 U. PhysiCal: A rapid calibration scan for B0, B1+, coil sensitivity and Eddy current mapping. In: *ISMRM*. 2020:0661.

59. Man LC, Pauly JM, Macovski A. Multifrequency interpolation for fast off-resonance correction. *Magn Reson Med*. 1997;37(5):785-792. doi:10.1002/mrm.1910370523
60. Le Bihan D, Mangin JF, Poupon C, et al. Diffusion tensor imaging: Concepts and applications. *Journal of Magnetic Resonance Imaging*. 2001;13(4):534-546. doi:10.1002/jmri.1076
61. Yang G, McNab JA. Eddy current nulled constrained optimization of isotropic diffusion encoding gradient waveforms. *Magn Reson Med*. 2019;81(3):1818-1832. doi:10.1002/MRM.27539
62. Aliotta E, Moulin K, Ennis DB. Eddy current–nulled convex optimized diffusion encoding (EN-CODE) for distortion-free diffusion tensor imaging with short echo times. *Magn Reson Med*. 2018;79(2):663-672.
63. Aliotta E, Wu HH, Ennis DB. Convex optimized diffusion encoding (CODE) gradient waveforms for minimum echo time and bulk motion–compensated diffusion-weighted MRI. *Magn Reson Med*. 2017;77(2):717-729. doi:10.1002/MRM.26166
64. Alsop DC. Phase insensitive preparation of single-shot RARE: Application to diffusion imaging in humans. *Magn Reson Med*. 1997;38(4):527-533. doi:10.1002/mrm.1910380404
65. Zhang Q, Coolen BF, Versluis MJ, Strijkers GJ, Nederveen AJ. Diffusion-prepared stimulated-echo turbo spin echo (DPsti-TSE): An eddy current-insensitive sequence for three-dimensional high-resolution and undistorted diffusion-weighted imaging. *NMR Biomed*. 2017;30(7). doi:10.1002/nbm.3719
66. Li H, Zu T, Chen R, et al. 3D diffusion MRI with twin navigator-based GRASE and comparison with 2D EPI for tractography in the human brain. *Magn Reson Med*. 2023;90(5):1969-1978. doi:10.1002/MRM.29769
67. Foo TKF, Laskaris E, Vermilyea M, et al. Lightweight, compact, and high-performance 3T MR system for imaging the brain and extremities. *Magn Reson Med*. 2018;80(5):2232-2245. doi:10.1002/mrm.27175
68. Keil B, Blau JN, Biber S, Hoecht P, Tountcheva V, Setsompop K, Triantafyllou C, Wald LL. A 64-channel 3T array coil for accelerated brain MRI. *Magn Reson Med*. 2013; 70(1):248-58. doi:10.1002/mrm.24427
69. Middione MJ, Loecher M, Cao X, Setsompop K, Ennis DB. Pre-excitation gradients for eddy current nulled convex optimized diffusion encoding (Pre-ENCODE). *Magn Reson Med*. 2024;92: 573-585.
70. Drago JM, Guerin B, Stockmann JP, Wald LL. Multiphoton parallel transmission (MP-pTx): Pulse design methods and numerical validation. *Magn Reson Med*. 2024; doi: 10.1002/mrm.30116
71. Liao C, Stockmann J, Cao X, Li Z, Craven-Brightman L, Sliwiak M, Biggs C, Zhong Z, Wang N, Wu H, Grafendorfer T, Robb F, Gruber B, Mareyam A, Kerr AB, Setsompop K. Flexible Use Of AC/DC Coil For Eddy-Currents And Concomitant Fields Mitigation With Applications In Diffusion-Prepared Non-Cartesian Sampling. ISMRM, 2023,program number: 1236.
72. Ramos-Llordén G, Lee HH, Davids M, Dietz P, Krug A, Kirsch JE, Mahmutovic M, Müller A, Ma Y, Lee H, Maffei C. Ultra-high gradient connectomics and microstructure MRI scanner for imaging of human brain circuits across scales. *Nature biomedical engineering.* 2025; doi: 10.1038/s41551-025-01457-x
73. Roemer PB and Hickey JS. Self-shielded gradient coils for nuclear magnetic resonance imaging. *US patent* 4737716, 1988.
74. Chapman BL, Doyle M, Pohost G. The design of optimized self-shielded electromagnetic coils of restricted length Soc. In Magn. Reson. Med. 12th Annu. Meeting 1992 (p. 585).
75. Wu D, Liu D, Hsu YC, Li H, Sun Y, Qin Q, Zhang Y. Diffusion-prepared 3D gradient spin-echo sequence for improved oscillating gradient diffusion MRI. *Magn Reson Med*. 2021; 85(1):78-88. doi: 10.1002/mrm.28401
76. Ajala A, Foo TK, Lee SK. Prospective compensation of second-order concomitant fields in a high-performance gradient system using a second-order harmonic shim coil. *Magn Reson Med*. 2025; doi: 10.1002/mrm.70094
77. Nguyen C, Fan Z, Sharif B, He Y, Dharmakumar R, Berman DS, Li D. In vivo three-dimensional high resolution cardiac diffusion-weighted MRI: a motion compensated diffusion-prepared balanced steady-state free precession approach. *Magn Reson Med*. 2014;72(5):1257-67. doi: 10.1002/mrm.25038

**Figure Caption**

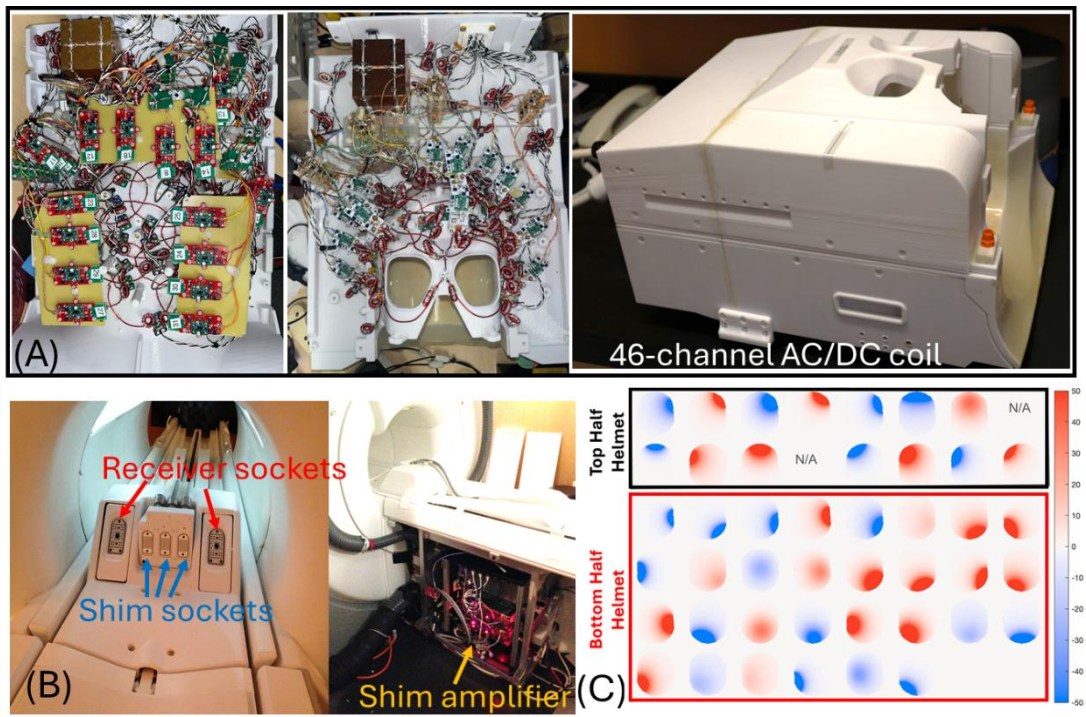

**Figure 1.** (A) Top and bottom half of the 46-channel AC/DC shim-array. (B) Receiver and shim sockets of the coil and shim amplifiers installed seamlessly into the space behind the MRI scanner. (C) An acquired calibration B0 map basis set shown on a single slice. 44 active shim channels are scanned in (C).

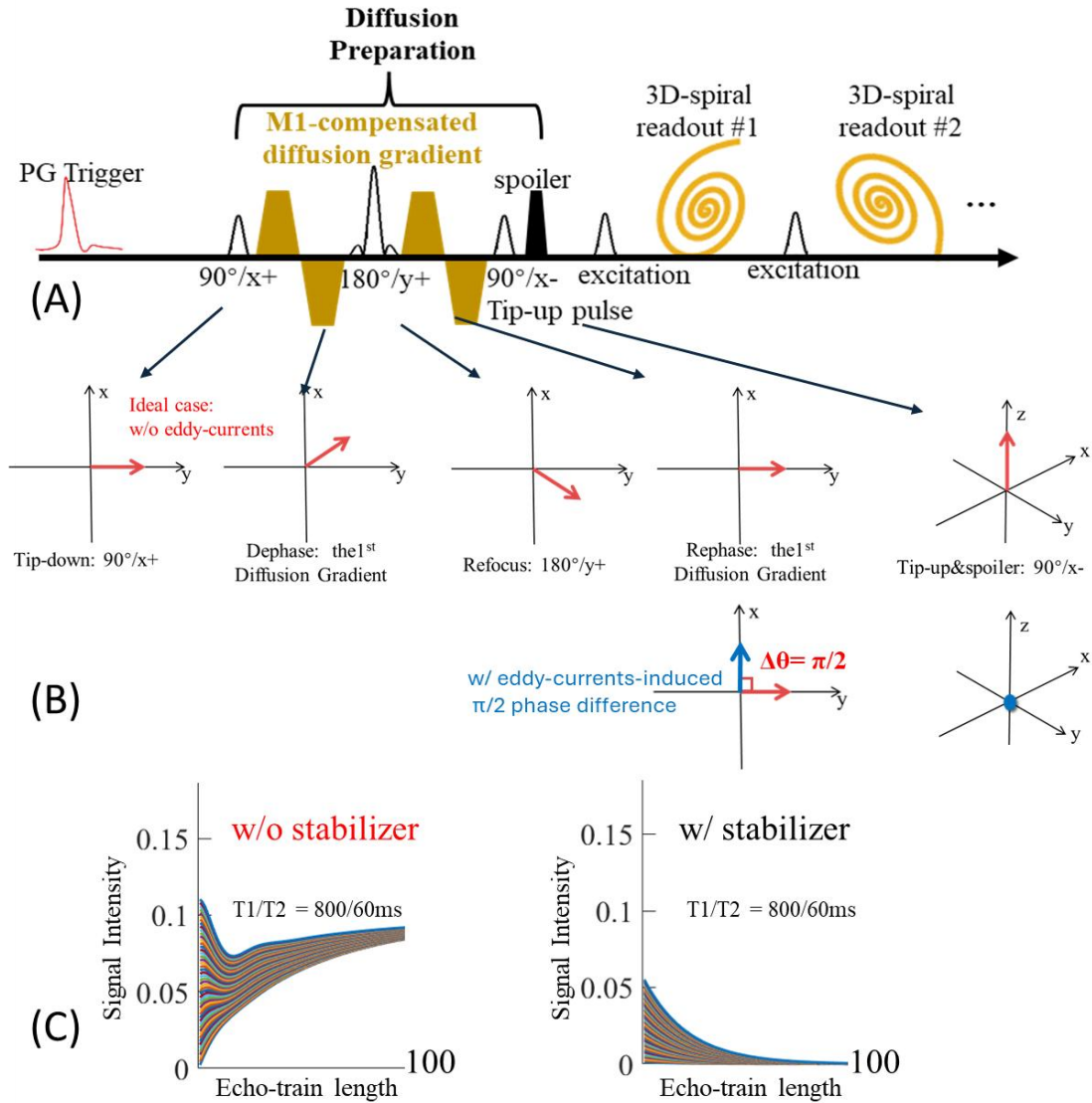

**Figure 2.** (A) Sequence diagram of a typical diffusion-prepared sequence. (B) Demonstration of signal loss in diffusion-prepared acquisition due to phase differences. In this example, the eddy-current-induced phase error is π/2, causing the magnetization affected by eddy currents to remain in the transverse (Mxy) plane rather than being tipped correctly to Mz, which leads to signal dropout in the resulting DP images. (C) signal evaluations with and without magnitude stabilizer. Given T1/T2 = 800/60 ms, Mxy of the 1st TR = 0.0816 without magnitude stabilizer while the Mxy of the 1st TR =0.0408 with magnitude stabilizer.

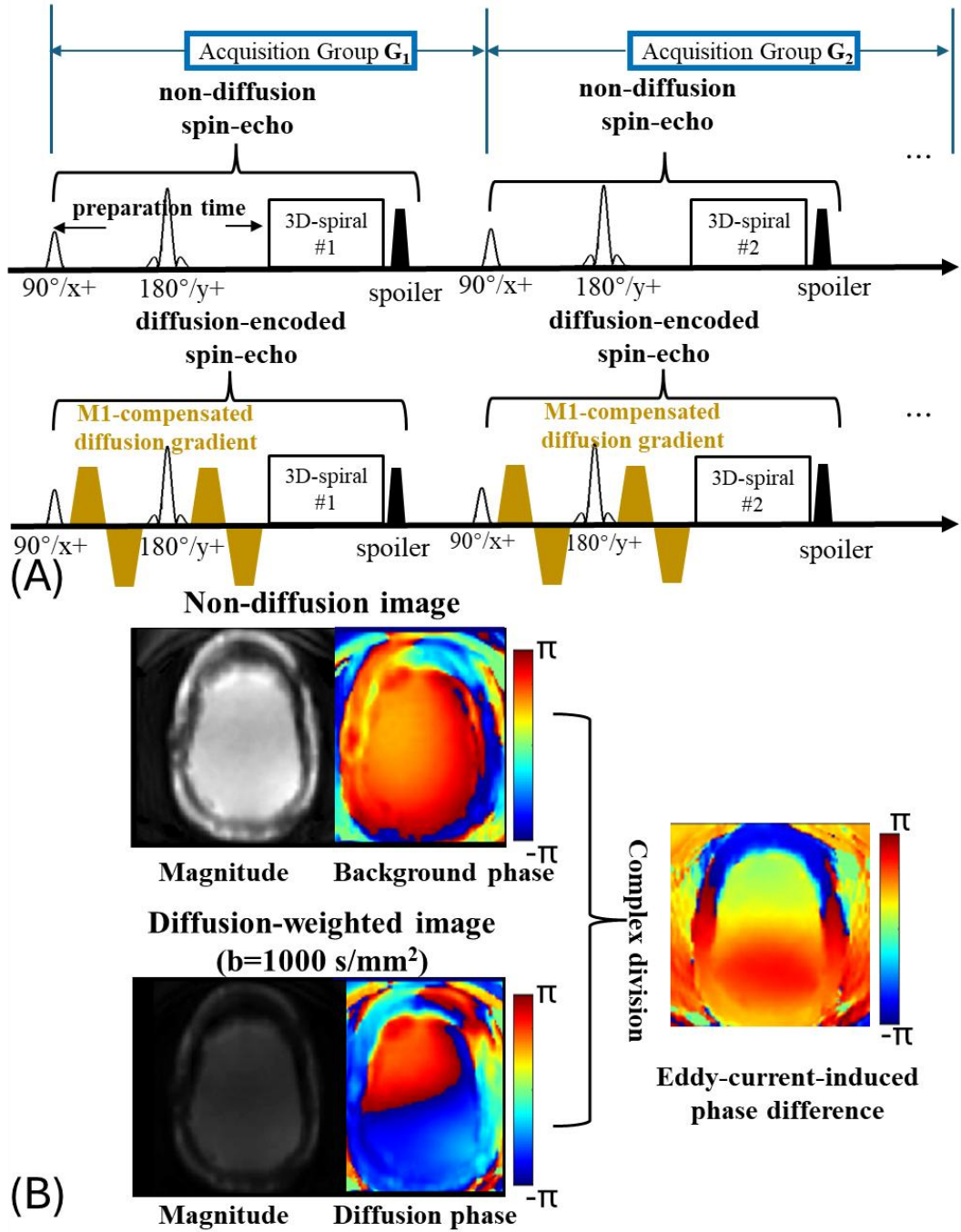

**Figure 3.** (A) A prescan spin-echo sequence to measure eddy current-induced phase differences. (B) The phase differences can be extracted by complex division between diffusion-weighted and non-diffusion images.

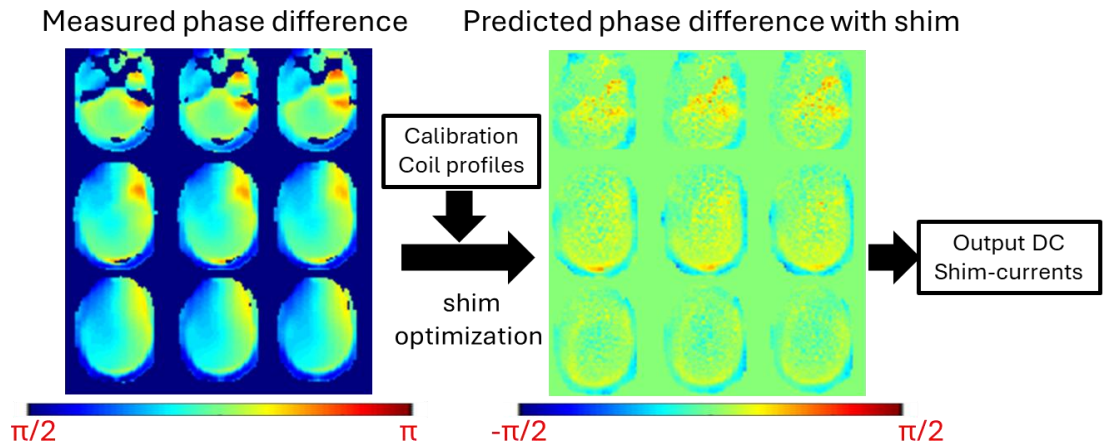

**Figure 4.** The measured phase differences were set as the input of the shim optimization, to create opposite phase variations to compensate the eddy current-induced phase differences.

*In vivo* 3D DP-FISP imaging with the AC/DC shim-array

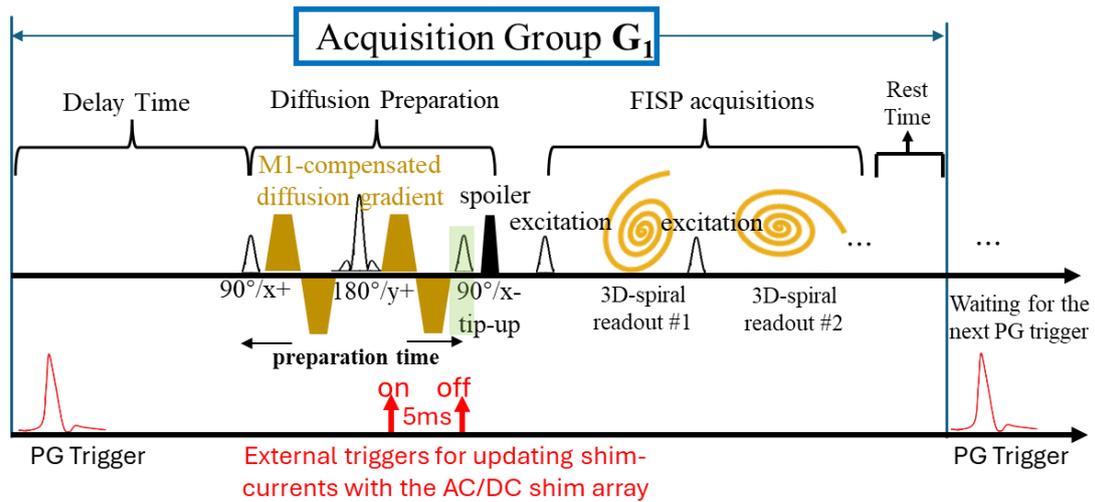

**Figure 5.** Cardiac-gated, M1-compensated diffusion-prepared multi-shot 3D FISP sequnece with the AC/DC shim-array. Tailored shim currents were applied during a 5ms-duration interval marked by external triggers, to correct for eddy current-induced phase error prior to the tip-up RF pulse. A pulse oximeter-gated (PG) trigger was employed to counteract the phase instabilities caused by cardiac pulsations during the diffusion-preparation.

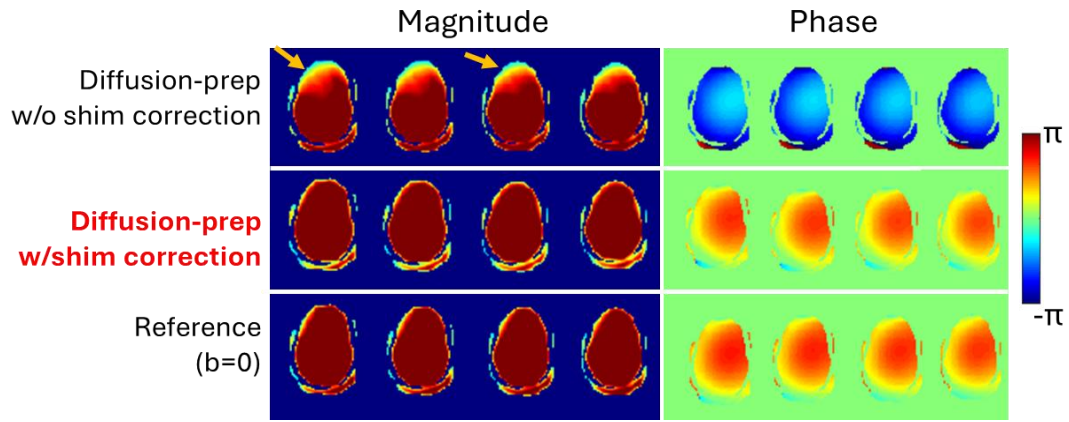

**Figure 6.** Diffusion-prepared (DP) images with and without shim correction. Compared to reference b=0 images, the DP image with shim-correction avoids the undesirable signal loss from eddy current-induced phase error, which is present in the DP images without correction (yellow arrows). With shim-correction, the phase maps of the DP images are similar to reference phase maps.

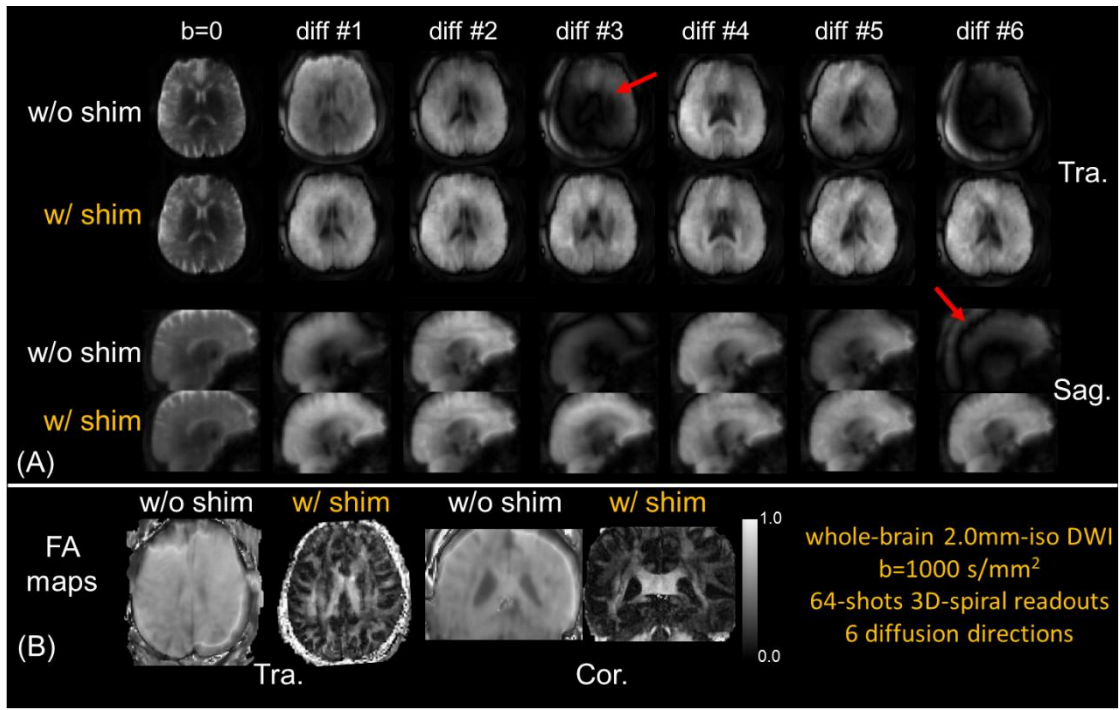

**Figure 7.** *In vivo* whole-brain six-diffusion-direction diffusion-prepared (DP) images with 2mm isotropic resolution (A) and corresponding fractional anisotropy (FA) maps (B) with and without shim correction at b=1000 s/mm$^2$. Red arrows indicate regions of signal dropout that are recovered with both pre-pulse and shim corrections.

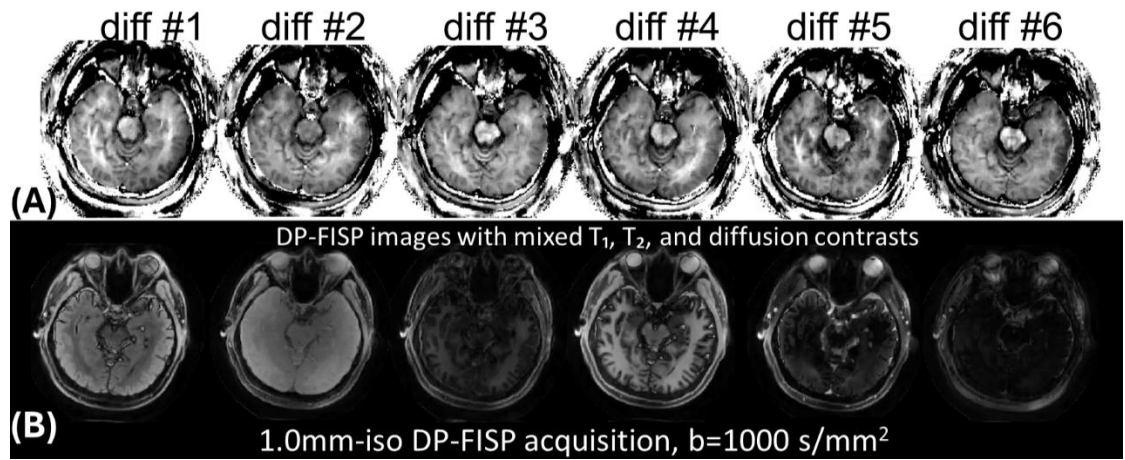

**Figure 8.** A representative bottom slice from an *in vivo* whole-brain diffusion-prepared (DP) acquisition with 1 mm isotropic resolution and six diffusion directions at b = 1000 s/mm$^2$, acquired using DP-FISP with shim correction. (A) Images from the six diffusion directions. (B) Selected time points following the DP module, illustrating the composite contrast from mixed $T_1$, $T_2$, and diffusion weighting achieved with the DP-FISP sequence.

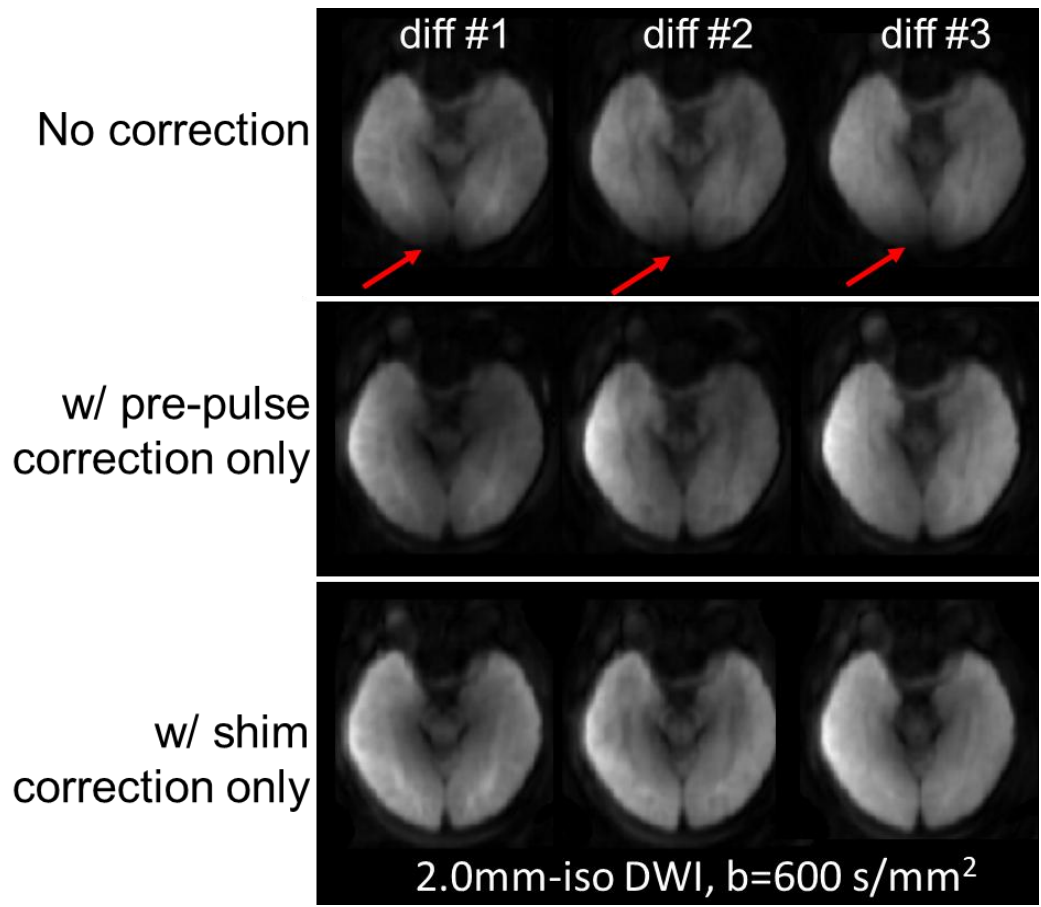

**Figure 9.** Comparison of diffusion images without correction, with pre-pulse correction, and with shim correction at b = 600 s/mm² for three diffusion directions. As indicated by the red arrows, both pre-pulse and shim corrections recovered signal dropout in the brain.

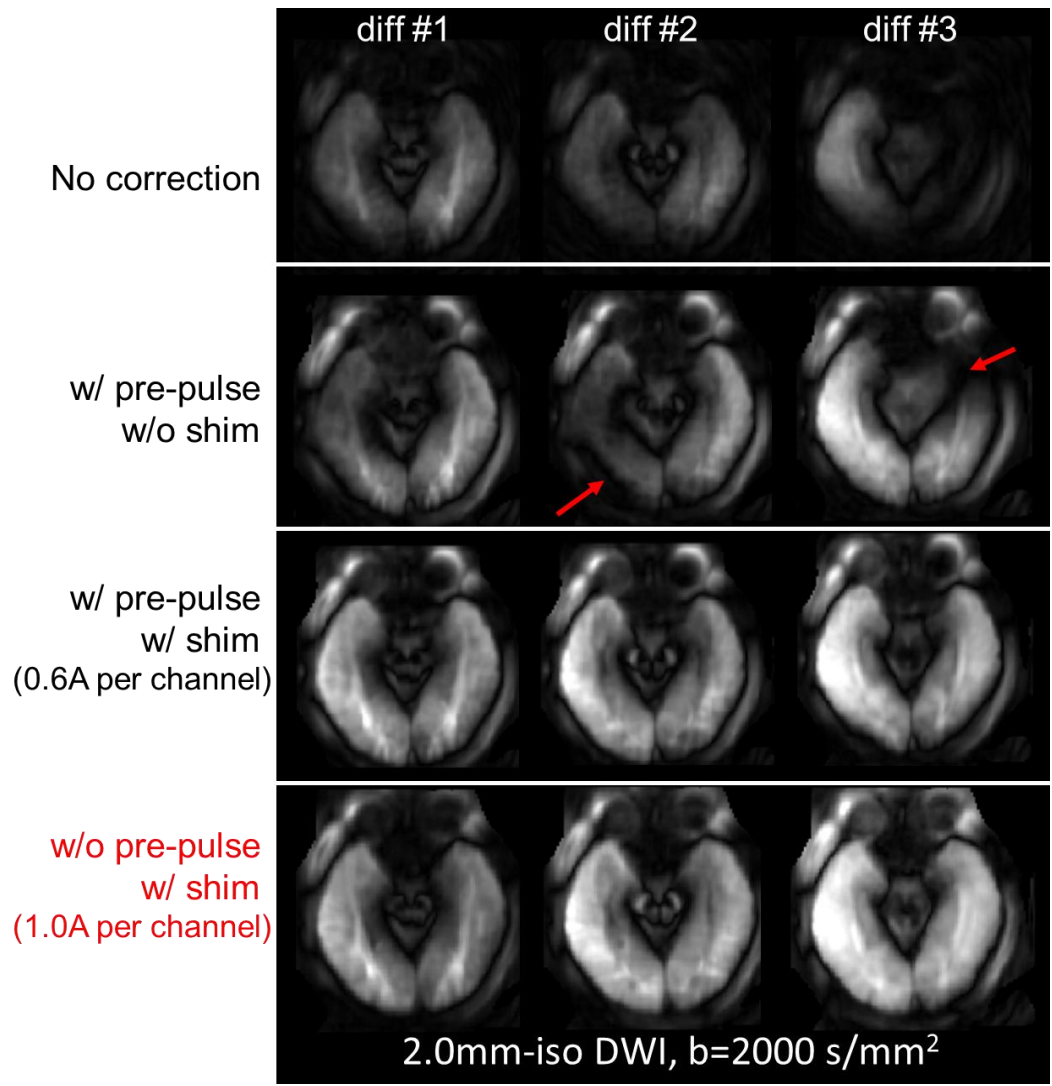

**Figure 10.** Diffusion-prepared (DP) images with and without pre-pulse correction and with and without shim correction.

# Supporting Information for "A dynamic shim approach for correcting eddy current effects in diffusion-prepared MRI acquisition using a multi-coil AC/DC shim-array" by Liao et al.

**S1 Linear approximation of eddy-current-induced phase using orthogonal gradient basis**

It is important to reduce the need for complete calibration for every new diffusion protocol. If the diffusion mixing time is kept constant, the eddy currents from any arbitrary diffusion direction could be expressed as a linear combination of the eddy currents measured along the three orthogonal gradient axes (Gx, Gy, and Gz). Specifically, we observe that the eddy currents scales as the square of the slew rate. For instance, for the same b-value and mixing time, a diffusion direction of $[1/\sqrt{3}, 1/\sqrt{3}, 1/\sqrt{3}]$ will require only 1/3 of the maximum Gx amplitude and slew rate compared to the [1, 0, 0] direction. As a result, the eddy current contribution from Gx in the former case would be $1/\sqrt{3}$ times smaller.

Relying on this weighting law, the following figure shows an example for the diffusion direction $[-0.487, 0.446, 0.75]$. The left panel displays the eddy-current-induced phase evolution directly measured from this diffusion direction, while the middle panel shows the simulated phase evolution based on a weighted linear combination of the Gx, Gy, and Gz basis measurements. The actual phase is well predicted by this linear combination, confirming that the eddy-current-induced phase along arbitrary diffusion directions can be accurately estimated from the orthogonal components. Minor deviations remain, as shown in the right panel (error map ×10), but are within ±5° variations, corresponding to less than 1% signal reduction in tip-up efficiency.

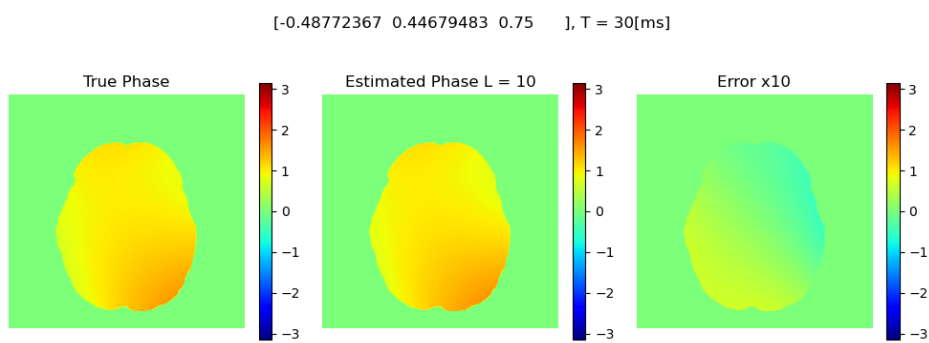

**Supporting Figure S1.** Comparison of linearly combined and measured eddy-current-induced phase.

**S2 Concomitant fields correction using AC/DC shim-array in non-Cartesian sampling**

Concomitant fields arising from gradient encoding can introduce additional phase accumulation during the readout. From Maxwell's equations, the concomitant field $B_c$ for symmetric gradient hardware is approximated to the second order of Taylor series expansion of magnet field **B**, which can be expressed as:

$$B_c(t) \approx \left(\frac{Gz(t)^2}{8B_0}\right)(X^2 + Y^2) + \left(\frac{Gx(t)^2 + Gy(t)^2}{2B_0}\right)Z^2 - \left(\frac{Gx(t)Gz(t)}{2B_0}\right)XZ - \left(\frac{Gy(t)Gz(t)}{2B_0}\right)YZ \quad (1)$$

where $B_0$ is the main magnetic field, (Gx(t), Gy(t), Gz(t)) are the time-dependent linear gradient fields of three axes and (X, Y, Z) are 3D spatial coordinates in the magnet. Using this equation, we can simulate the concomitant field for arbitrary trajectories at arbitrary positions at a given time-point, denoted as 't'. Equation 1 demonstrates that the concomitant field accumulates over acquisition time. This accumulated concomitant field introduces an additional phase during the readout, which in turn leads to image blurring. The image blurring becomes particularly pronounced when using high gradient strengths, extended readout durations, and large offsets from the imaging isocenter.

To address the blurring caused by concomitant fields, we propose a dynamic approach for compensating the additional phase using the AC/DC shim array. This approach consists of the following steps: (i) Given the gradient waveforms of the acquisition trajectories and slice position, we simulate concomitant fields across time using Equation 1. To account for the temporal switch limits of the AC/DC shim array, the temporal resolution of the simulation is set to 2ms. (ii) The trajectories are segmented at intervals of 2ms, and the accumulated phase per segment is determined by integrating the concomitant fields over the Δt = 2ms segment, expressed as: $2\pi \int_t^{t+\Delta t} \mathbf{B}_c(t)dt$. (iii) The phase increments obtained for all segments are then fed into a shim optimization algorithm to minimize them to zero. The shim optimization was calculated to correct concomitant field only in the phantom or brain region to make the optimization problem easier to solve. Similar to eddy current correction, the resulting optimal shim currents could then be updated every 2ms upon receiving external triggers. This updating process generates opposite phase maps for each segment, effectively compensating for the concomitant fields segment by segment during the readout.

The simulation procedure was as follows:

(1) We used a clean phantom image as the target. To obtain the reference reconstructed image with no concomitant field contamination, we performed a non-uniform FFT (NUFFT) plus coil combined reconstruction on a clean multi-channel k-space data obtained by Fourier transforming coil sensitivity weighted phantom images into k-space along the k-space trajectory of a 64 ms single-shot 2D spiral trajectory. The trajectory design parameters were FOV = 220 mm, maximum gradient strength Gmax = 83.8 mT/m, maximum slew rate = 600 T/m/s, and in-plane resolution = 1mm, as shown in the following figure:

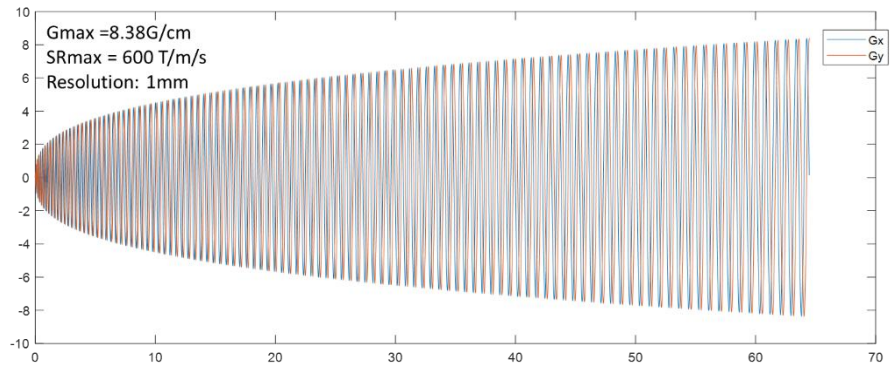

(2) The concomitant fields were calculated according to Equation 1. For our simulation, a high temporal resolution of 1μs was used to accurately capture the continuous phase accrual, while the applied AC/DC coil eddy current mitigation was performed at a 2ms temporal increment, with the value of the correction field hold constant in each 2ms period. Simulations were performed at a rotated slice with Z = +40 mm and a double-oblique rotation of 10° and 20°. The following figure shows the simulated phase accrual induced by the concomitant fields

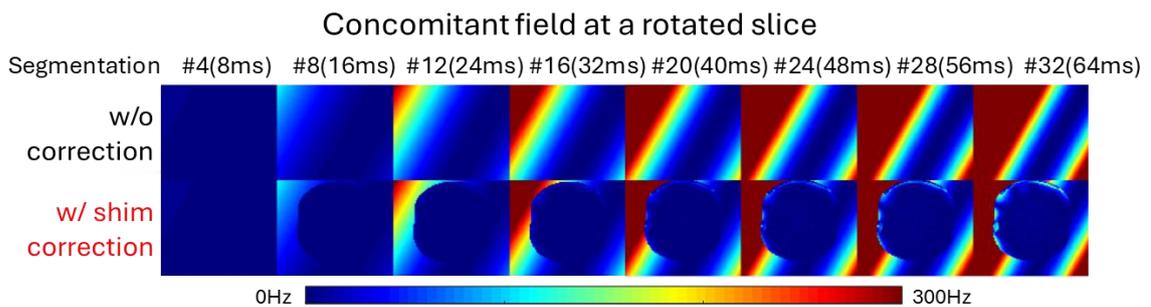

(3) To simulate k-space data that is corrupted with concomitant fields, for each time point along the readout, the net phase accrual induced by the concomitant fields was added to the phantom image, followed by time-point-by-time-point NUFFT processing. The subsequent inverse NUFFT reconstruction resulted in image blurring due to the accumulated concomitant-field-induced phase, as shown below:

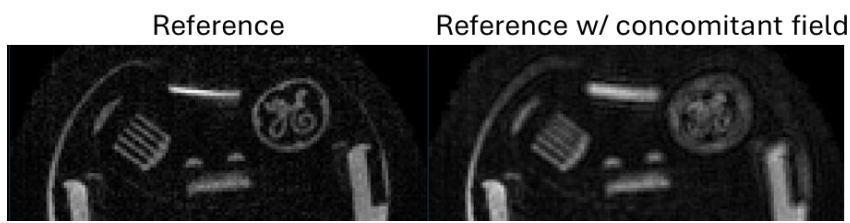

(4) To correct this effect, we simulate external triggers for dynamic shim updating with the 48-channel AC/DC shim array at 2ms interval across the spiral readout of 64 ms, corresponding to 32 updates. By generating opposing fields to cancel the concomitant-field-induced phase accrual, the corrected images obtained after inverse NUFFT reconstruction showed a substantial reduction of blurring artifacts, as depicted in the following figure:

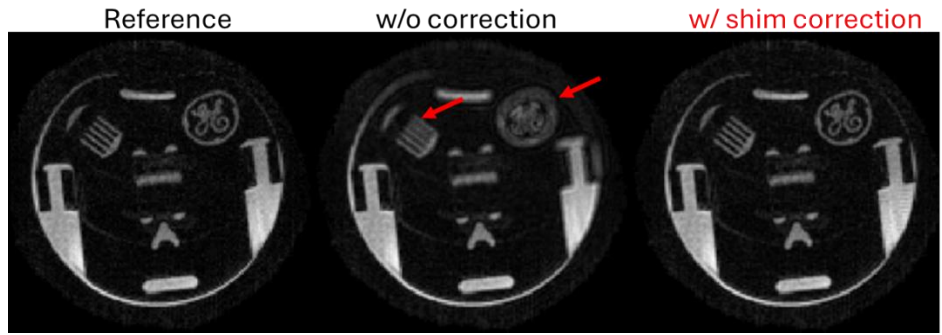

(5) To test whether increasing the update rate of the dynamic shim would enhance concomitant field correction, we also implemented a faster 0.5ms update. However, we found that it did not provide additional improvements over the current 2ms update rate. The simulation results are shown below:

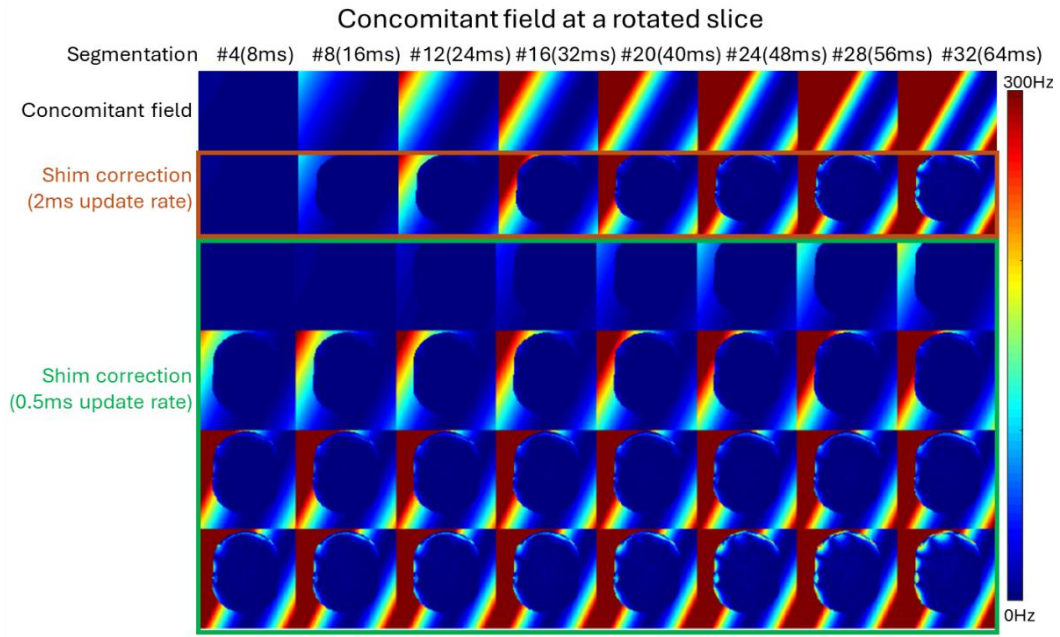